\documentclass[reprint, nofootinbib]{revtex4}
\usepackage[english]{babel}
\usepackage{slashed}
\usepackage{amsmath,graphicx,xcolor}

\catcode`\>=\active \def>{
\fontencoding{T1}\selectfont\symbol{62}\fontencoding{\encodingdefault}}
\newcommand{\nobracket}{}
\newcommand{\tmop}[1]{\ensuremath{\operatorname{#1}}}
\newcommand{\tmfloatcontents}{}
\newlength{\tmfloatwidth}
\newcommand{\tmfloat}[5]{
  \renewcommand{\tmfloatcontents}{#4}
  \setlength{\tmfloatwidth}{\widthof{\tmfloatcontents}+1in}
  \ifthenelse{\equal{#2}{small}}
    {\ifthenelse{\lengthtest{\tmfloatwidth > \linewidth}}
      {\setlength{\tmfloatwidth}{\linewidth}}{}}
    {\setlength{\tmfloatwidth}{\linewidth}}
  \begin{minipage}[#1]{\tmfloatwidth}
    \begin{center}
      \tmfloatcontents
      \captionof{#3}{#5}
    \end{center}
  \end{minipage}}

\begin{document}

\title{On the axial anomaly in Very Special Relativity}
\date{\today}

\author{
	Jorge Alfaro$^{a}$
}

\affiliation{
	$^{a}$Instituto de F\'{i}sica, Pontificia Universidad de Cat\'olica de Chile, \mbox{Avda. Vicu\~na Mackenna 4860, Santiago, Chile}
}

\begin{abstract} 

In this paper we study the axial anomaly in Very Special Relativity  Electrodynamics using
Pauli-Villars and dimensional regularization of ultraviolet divergences and
Mandelstam-Leibbrandt regularization of infrared divergences. We compute the
anomaly in 2 and 4 dimensional space-time. We find that this procedure
preserves the vector Ward identity(charge conservation) and reproduce the
standard axial anomaly in 2 and 4 dimensions without corrections from VSR.
Finally, we show how to obtain the anomaly in the path integral approach.
\end{abstract}

\maketitle 

\section{Introduction}

The Standard Model of Particle Physics(SM) is a very successful theory. With the
discovery of the Higgs boson at CERN,  its particle composition was
completed\cite{w2}.

But the discovery of neutrino oscillations showed that the neutrinos have mass
whereas in the SM they are massless\cite{Langacker}.

One of the most important problems of Particle Physics is to provide a mass
for the neutrino without disturbing the chiral nature of the SM, since  neutrinos
appear to be left handed.

The seesaw mechanism is a popular mechanism to obtain massive neutrinos\cite{mohapatra}. However, it means to introduce new particles and new interactions.

One possibility to have massive chiral neutrinos is Very Special
Relativity(VSR)\cite{CG1}

VSR assumes that the true symmetry of Nature is not the full Lorentz group,
but some of its subgroups. The most interesting of these subgoups are
$Sim ( 2 )$ and $Hom ( 2 )$. Using these subgroups new terms are allowed
such that the neutrino get a mass\cite{CG2}.

Some time ago, we  proposed the SM with VSR{\cite{ja1}} (VSRSM).Its particle composition and interactions are the same as in  the SM, but neutrinos can have a VSR mass
without lepton number violation.

Loop computations in VSR are non trivial though. New infrared divergences appear and they have to be regularized. We studied how to do so using the calculation of integrals in the Mandelstam-Leibbrandt (ML) prescription\cite{Mandelstam},\cite{Leibbrandt} introduced in \cite{alfaroML}, in \cite{japlb} and \cite{jauniverse}. The Ward identities corresponding to the gauge  and the $Sim(2)$ symmetry of the model are preserved.

Last year, we applied these techniques to the Schwinger model in VSR \cite{as} and to the photon mass in VSR \cite{as2}.

A very important test that the VSRSM has to pass is the cancellation of axial anomalies. Being a chiral local gauge symmetry model, the presence of chiral anomalies may kill the model, because the gauge symmetry will be lost and renormalizability and unitarity could not be simultaneously satisfied.

In  \cite{as} we did a computation of the two dimensional axial anomaly. We obtained that the vector current is conserved and the axial anomaly get a correction from VSR
in the form of a multiplicative factor.

The authors of \cite{alex} tried to compute the axial anomaly in four dimensions using the prescription to treat $\gamma^5$ introduced in \cite{mdr}.
They claim that there is an anomaly in the vector current as well as in the axial vector current. However their computation missed two important graphs. (Please see chapter IV). 

In this paper we study the axial anomaly in two and four dimensions using Pauli-Villars (PV) and dimensional(DR) regularization of ultraviolet divergences and ML prescription for infrared divergences. We show explicitly that the vector current is conserved and that the axial anomaly is the same we get in Lorentz invariant Electrodynamics, without any correction from VSR. Our result relies on two properties of the ML prescription: First, it allows shifting of the loop momentum variable(which implies gauge invariance) and second, it respects naive power counting.

According to this result, the VSRSM must be free from anomalies and therefore consistent.

The paper is written as follows. In chapter II we define the lagrangian of VSR Electrodynamics and derive the Feynman rules that will be used 
to compute the anomalies.In chapter III we compute the axial anomaly in two dimensional space time. In chapter IV we study the axial anomaly in four dimensions. In chapter V we study the axial anomaly in 2d using DR. In chapter VI, we derive the axial anomaly in 4d, using DR. In chapter VII we present the derivation of the axial anomaly using the path integral. Finally in chapter VIII we draw some conclusions.

\section{The model}

The Electrodynamics sector of the VSRSM in the Feynman gauge.
\begin{eqnarray}
  \mathcal{L} = \bar{\psi}  \left( i \left( \slashed{D} + \frac{1}{2} \slashed{n}
  m^{2} (n \cdot D)^{-1} \right) -M \right) \psi - \frac{1}{4} F_{\mu \nu}
  F^{\mu \nu} - \frac{( \partial_{\mu} A_{\mu} )^{2}}{4} &  &  \label{qed}\\
  D_{\mu} = \partial_{\mu} -ieA_{\mu} , & F_{\mu \nu} = \partial_{\mu} A_{\nu}
  - \partial_{\nu} A_{\mu} &  \nonumber
\end{eqnarray}

The vector current(electric charge conservation) is:
\[ j^{\mu} = \bar{\psi}   \gamma^{\mu}   \psi + \frac{1}{2} m^{2} \left(
   \frac{1}{n  \cdot D^{\dag}} \bar{\psi} \right) \slashed{n} n^{\mu} \left(  
   \frac{1}{n  \cdot D}   \psi \right) \]
The axial vector current is:
\[ j^{\mu 5} =  \bar{\psi}   \gamma^{\mu}   \gamma^{5} \psi + \frac{1}{2}
   m^{2} \left( \frac{1}{n  \cdot D^{\dag}} \bar{\psi} \right) \slashed{n} n^{\mu}
   \gamma^{5} \left(   \frac{1}{n  \cdot D}   \psi \right)   \]
Both currents are conserved at the classical level\cite{as}.
We are interested in computing expectation values of these currents.

To get the Feynman rules we use the expansion of $( n.D )^{-1}$ both in the
currents and the lagrangian.
\begin{eqnarray*}
  ( n.D )^{-1} = ( 1+i e ( n. \partial )^{-1} ( n.A ) + ( i e )^{2} ( n.
  \partial )^{-1} ( n.A ) ( n. \partial )^{-1} ( n.A ) + ( i e )^{3} ( n.
  \partial )^{-1} ( n.A ) ( n. \partial )^{-1} ( n.A ) ( n. \partial )^{-1} (
  n.A ) ) ( n. \partial )^{-1} &  & 
\end{eqnarray*}
The Feynman rules are listed in Appendix A.

\section{Two dimensional axial anomaly}
In this case we have to compute the expectation value of the axial vector
current in a background field $A_{\nu}$. We use the convention of \cite{Peskin},$\epsilon^{01}=+1$. 
\begin{equation}
	<j^{5\nu} ( q ) > =\int d^{2}x <j^{5\nu}( x ) >e^{i q x} = ( -i e
	)^{-1} i  \Pi^{5\mu \nu} ( q ) A_{\mu}
\end{equation}
The contribution to the two dimensional anomaly in VSR Electrodynamics is
given by the two graphs (Figure \ref{fig:2d1} and Figure \ref{fig:2d2}):
\begin{figure}[h!]
	\centering
	\includegraphics[scale=0.65]{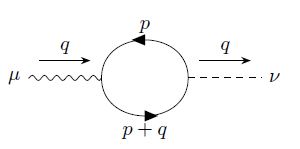}
	\caption{}
	\label{fig:2d1}
\end{figure}

\begin{eqnarray}
	i  \Pi^{ 1 5 \mu \nu} =- ( -i e )^{2} 
	\int d p  \tmop{Tr} \{
	\left[ \gamma^{\mu} + \frac{1}{2} n^{\mu} \left( \slashed{n} \right) m^{2}   (n. ( p+q ) )^{-1} ( n.p )^{-1} \right] \nonumber \\
	\frac{i \left( \slashed{p} +M-\frac{m^{2}}{2}  \frac{\slashed{n}}{n \cdot p} \right)}{p^{2} -M^{2} -m^{2} +i\varepsilon}
		 \left[ \gamma^{\nu} + \frac{1}{2} n^{\nu} \left( \slashed{n}
	\right) m^{2}   ( n. ( p+q ) )^{-1} ( n.p )^{-1} \right]
	 \gamma^{5} 
	 \frac{i \left( \left( \slashed{p} + \slashed{q} \right) +M- \frac{m^{2}}{2} 
		\frac{\slashed{n}}{n \cdot ( p+q )} \right)}{( p+q )^{2} -M^{2} -m^{2} +i
		\varepsilon} \} \label{2d1}
\end{eqnarray}
\begin{figure}[h!]
	\centering
	\includegraphics[scale=0.65]{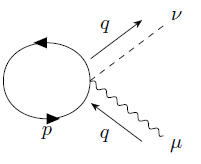}
	\caption{}
	\label{fig:2d2}
\end{figure}

\begin{eqnarray}
  i  \Pi^{2 5\mu \nu} = ( -1 ) ( i e )^{2} n^{\mu} n^{\nu} i \int  d p (
  n.p )^{-1} ( n.p )^{-1} [ ( n. ( q+p ) )^{-1} + ( n. ( -q+p ) )^{-1} ]
  \tmop{Tr} \{\frac{1}{2} \slashed{n} m^{2}  \frac{i \left( \slashed{p} +M-
  \frac{m^{2}}{2}  \frac{\slashed{n}}{n \cdot p} \right)}{p^{2} -M^{2} -m^{2} +i
  \varepsilon} \gamma^{5}\} &  &  \label{2d2}
\end{eqnarray}

To compute the axial anomaly we will use Pauli-Villars regularization and
Mandelstam-Leibbrandt prescription to treat infrared divergences. We will
follow reference \cite{Pokorski}.

Notice that equation (\ref{2d1}) is logarithmically divergent and equation
(\ref{2d2}) is finite.

It is easy to check that formally:
\[ q_{\mu} ( \Pi^{1 5\mu \nu} + \Pi^{2 5\mu \nu}) =0 \]
if shift of the integration variable $p \rightarrow p+k$ is allowed. Here $k$
is a constant vector. This would be true if the integral (\ref{2d1}) would be
finite.

Introduce a Pauli-Villars particle of mass $\bar{M}$ and define the
regularized amplitude:
\begin{eqnarray*}
  \Pi^{5R\mu \nu} ( M, \bar{M} ,q ) = \Pi^{1 5 \mu \nu} ( M,q ) + \Pi^{2 5
  \mu \nu} ( M,q ) - \Pi^{1 5\mu \nu} ( \bar{M} ,q ) - \Pi^{2 5\mu
  \nu}( \bar{M} ,q ) &  & 
\end{eqnarray*}
Since $\Pi^{5R\mu \nu} ( M, \bar{M} ,q )$ is finite, it satisfies the naive
Ward identity(electric charge conservation):
\[ q_{\mu} \Pi^{5R\mu \nu} ( M, \bar{M} ,q ) =0 \]
On the other hand, the axial Ward identity is, formally:
\[ i ( \Pi^{1 5\mu \nu} + \Pi^{2 5\mu \nu} ) q_{\nu} =2M\mathcal{A} (
   M,q )^{\mu} \]
\begin{equation}
  \begin{array}{lll}
    =2M ( -i e )^{2} \int d p  \tmop{Tr} \left\{ \left[ \gamma^{\mu} +
    \frac{1}{2} n^{\mu} \left( \slashed{n} \right) m^{2}   ( n. ( p+q ) )^{-1} (
    n.p )^{-1} \right]  \frac{i \left( \slashed{p} +M- \frac{m^{2}}{2} 
    \frac{\slashed{n}}{n \cdot p} \right)}{p^{2} -M^{2} -m^{2} +i \varepsilon}
    \gamma^{5} \frac{i \left( \left( \slashed{p} + \slashed{q} \right) +M-
    \frac{m^{2}}{2}  \frac{\slashed{n}}{n \cdot ( p+q )} \right)}{( p+q )^{2}
    -M^{2} -m^{2} +i \varepsilon} \right\} \label{2dano} &  & 
  \end{array}
\end{equation}
if shift of the integration variable $p \rightarrow p+k$ is allowed.

Therefore the regularized amplitude satisfies:
\[ i \Pi^{5R\mu \nu}( M, \bar{M} ,q ) q_{\nu} =2M\mathcal{A} ( M,q )^{\mu}
   -2 \bar{M} \mathcal{A} ( \bar{M} ,q )^{\mu} \]
Since the original amplitude is obtained formally as $l i m_{\bar{M}
\rightarrow \infty}$, the axial anomaly is given by:
\[ B^{\mu} =l i m_{\bar{M} \rightarrow \infty} ( -2 \bar{M} \mathcal{A} (
   \bar{M} ,q )^{\mu} ) \]
Now, we compute (\ref{2dano}). First notice that after computing the trace,
the integral is finite. A tipical term containing the vector $n^{\mu}$ is of
the form:
\begin{eqnarray*}
  C^{\mu} =2M^{2} m^{2} ( -i e )^{2} \varepsilon^{\mu \alpha} n_{\alpha} \int
  d p \frac{1}{p^{2} -M^{2} -m^{2} +i \varepsilon} \frac{1}{( p+q )^{2} -M^{2}
  -m^{2} +i \varepsilon} \frac{1}{n.p} &  & 
\end{eqnarray*}
Now we recall an important property of ML prescription. It preserves naive
power counting. According to this, $C^{\mu} \sim M^{-1}$ for large $M$.

Following the same argument, we can easily check that all terms containing
$n^{\mu}$ vanish when $M \rightarrow \infty$.

It remains the Lorentz invariant term:
\begin{eqnarray}
	i  \Pi^{5\mu \nu} ( q ) q_{\nu} =l i m_{\bar{M} \rightarrow \infty} &  & 
	-4e^{2} \bar{M}^2 \varepsilon^{\alpha \mu} q_{\alpha} \int d p
	\frac{1}{p^{2} - \bar{M}^{2} -m^{2} +i \varepsilon} \frac{1}{( p+q )^{2} -
		\bar{M}^{2} -m^{2} +i \varepsilon} = 
	-i\frac{e^{2}}{\pi} \varepsilon^{\alpha \mu} q_{\alpha}
\end{eqnarray}
\begin{equation}
 q_{\nu} <j^{5\nu} > = \frac{1}{-i e} i  \Pi^{5 \mu \nu} ( q )
A_{\mu} q_{\nu}  = \frac{e}{\pi}
\varepsilon^{\alpha \mu} q_{\alpha} A_{\mu} \label{2danomaly}
\end{equation}
Equation (\ref{2danomaly}) is the standard Lorentz invariant result\cite{Peskin}.

We want to comment on a previous computation of the anomaly in \cite{as}.
There and here, the vector current is conserved, but a different axial anomaly
is obtained. This difference may be a result of different normalization
conditions\cite{Pokorski} or the extra freedom we have when Lorentz symmetry is broken\cite{aas} .

It is clear though that the procedure used in \cite{as} does not respect naive power counting of the loop integrals.

\section{Four dimensional axial anomaly}

We compute:
\[ \int d^{4} x e^{-i r x} <p,q | j^{\mu 5} ( x ) | 0> = ( 2 \pi )^{4} \delta
( -r+p+q ) \varepsilon^{\ast}_{\nu} ( q ) \varepsilon^{\ast}_{\delta} ( p )
i  \Pi^{\mu \nu \delta} \]

There are four graphs that contribute to the axial anomaly in four dimensions
(Figure 3-6). Notice that in \cite{alex} Figure 5,6 are missing. They are fundamental
to satisfy the Ward identity for the vector current(charge conservation) as
well as the right computation of the axial anomaly.

\begin{figure}[h!]
	\centering
	\includegraphics[scale=0.65]{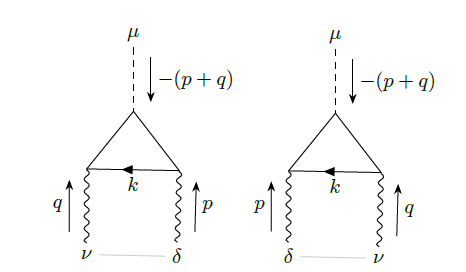}
	\caption{$\Pi^{ 5\mu \nu \delta}$}
	\label{fig:4d1}
\end{figure}

\begin{eqnarray}
  i  \Pi^{1 5\mu \nu \delta} =- ( -i e )^{2} \int d k  \tmop{Tr}
  \{ \left[ \gamma^{\mu} + \frac{1}{2} n^{\mu} \left( \slashed{n} \right)
  m^{2}   ( n. ( k+q ) )^{-1} ( n. ( k  - p ) )^{-1} \right]
  \gamma^{5}  \frac{i \left( \left( \slashed{k} + \slashed{q} \right) +M-
  \frac{m^{2}}{2}  \frac{\slashed{n}}{n \cdot ( k+q )} \right)}{( k+q )^{2} -M^{2}
  -m^{2} +i \varepsilon} \nonumber \\
\left[ \gamma^{\nu} + \frac{1}{2} n^{\nu} \left(
  \slashed{n} \right) m^{2}   ( n. ( k+q ) )^{-1} ( n.k )^{-1} \right] \frac{i
  \left( \slashed{k} +M- \frac{m^{2}}{2}  \frac{\slashed{n}}{n \cdot k} \right)}{k^{2}
  -M^{2} -m^{2} +i \varepsilon} \nonumber \\
\left[ \gamma^{\delta} + \frac{1}{2}
  n^{\delta} \left( \slashed{n} \right) m^{2}   ( n. ( k-p ) )^{-1} ( n.k )^{-1}
  \right] \frac{i \left( \slashed{k} - \slashed{p} +M- \frac{m^{2}}{2} 
  \frac{\slashed{n}}{n \cdot ( k-p )} \right)}{( k-p )^{2} -M^{2} -m^{2} +i
  \varepsilon} \} + ( p, \delta ) \rightarrow ( q, \nu ) &  & 
  \label{4d1}
\end{eqnarray}
\begin{figure}[h!]
	\centering
	\includegraphics[scale=0.65]{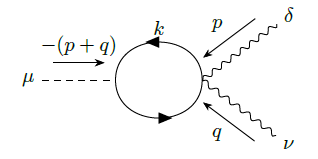}
	\caption{$\Pi^{2 5\mu \nu \delta}$}
	\label{fig:4d2}
\end{figure}

\begin{eqnarray}
  i  \Pi^{2 5\mu \nu \delta} = ( -1 ) ( i e )^{2} n^{\delta} n^{\nu} i \int
  d k ( n.k )^{-1} ( n. ( k-p-q ) )^{-1} [ ( n. ( k-q ) )^{-1} + ( n. ( k-p )
  )^{-1} ] \nonumber \\
  \tmop{Tr} \{\frac{1}{2} \slashed{n} m^{2}  \frac{i \left( \slashed{k} -
  \slashed{p} - \slashed{q} +M- \frac{m^{2}}{2}  \frac{\slashed{n}}{n \cdot ( k-p-q )}
  \right)}{( k-p-q )^{2} -M^{2} -m^{2} +i \varepsilon} \left[ \gamma^{\mu} +
  \frac{1}{2} n^{\mu} \left( \slashed{n} \right) m^{2}   ( n.k )^{-1} ( n. ( k-p-q
  ) )^{-1} \right] \gamma^{5} &  &  \nonumber\\
  \frac{i \left( \slashed{k} +M- \frac{m^{2}}{2}  \frac{\slashed{n}}{n \cdot k}
  \right)}{k^{2} -M^{2} -m^{2} +i \varepsilon} \} \nobracket &  &  \label{4d2}
\end{eqnarray}
\begin{figure}[h!]
	\centering
	\includegraphics[scale=0.65]{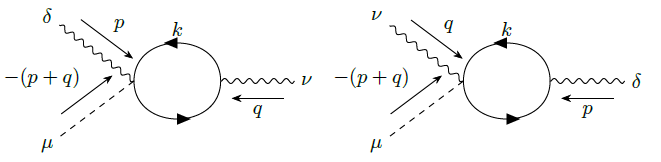}
	\caption{$\Pi^{3 5\mu \nu \delta}$}
	\label{fig:4d3}
\end{figure}

\begin{eqnarray}
  i  \Pi^{3 5\mu \nu \delta}= ( -1 ) ( i e )^{2} n^{\delta} n^{\mu} i \int
  d k ( n.k )^{-1} ( n. ( k-q ) )^{-1} [ ( n. ( k {\color{red} -} q
  {\color{red} -} p ) )^{-1} + ( n. ( k+p ) )^{-1} ] \nonumber \\
  \tmop{Tr} \left\{
  \frac{1}{2} \slashed{n} m^{2} \gamma^{5}  \frac{i \left( \slashed{k} +M-
  \frac{m^{2}}{2}  \frac{\slashed{n}}{n \cdot k} \right)}{k^{2} -M^{2} -m^{2} +i
  \varepsilon} \left[ \gamma^{\nu} + \frac{1}{2} n^{\nu} \left( \slashed{n}
  \right) m^{2}   ( n.k )^{-1} ( n. ( k-q ) )^{-1} \right] \frac{i \left(
  \slashed{k} - \slashed{q} +M- \frac{m^{2}}{2}  \frac{\slashed{n}}{n \cdot ( k-q )}
  \right)}{( k-q )^{2} -M^{2} -m^{2} +i \varepsilon} \right\}\nonumber \\
 + ( p, \delta )
  \rightarrow ( q, \nu ) &  &  \label{4d3}
\end{eqnarray}
\begin{figure}[h!]
	\centering
	\includegraphics[scale=0.65]{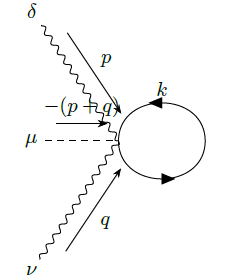}
	\caption{$\Pi^{4 5\mu \nu \delta} $}
	\label{fig:4d4}
\end{figure}

\begin{eqnarray}
	i  \Pi^{4 5\mu \nu \delta} =
  ( - 1 ) ( i e )^{2} n^{\nu} n^{\mu} n^{\delta} i \int  d
  k \{ \frac{1}{n.k} \frac{1}{n.k} 
  [ \frac{1}{n. ( k+p+q )}
  \frac{1}{n. ( k+p )} + \frac{1}{n. ( k+p+q )} \frac{1}{n. ( k+q )} +\nonumber\\
  \frac{1}{n. ( k-p )} \frac{1}{n. ( k-p-q )} + \frac{1}{n. ( k-q )}
  \frac{1}{n. ( k+p )} + \frac{1}{n. ( k-p )} \frac{1}{n. ( k+q )} +
  \frac{1}{n. ( k-q )} \frac{1}{n. ( k-p-q )} ] \} \nonumber\\
  \tmop{Tr}
  \left\{ \frac{1}{2} \slashed{n} m^{2} \gamma^{5}  [ i ] \frac{i \left( \slashed{k}
  +M- \frac{m^{2}}{2}  \frac{\slashed{n}}{n \cdot ( k )} \right)}{( k )^{2} -M^{2}
  -m^{2} +i \varepsilon} \right\} &  &  \label{4d4}
\end{eqnarray}

Notice that $\Pi^{2 5\mu \nu \delta} , \Pi^{3 5\mu \nu \delta} ,
\Pi^{4 5\mu \nu \delta}$ are ultraviolet finite. Only $\Pi^{1 5\mu \nu
\delta}$ is linearly divergent as in the Lorentz invariant electrodynamics.

To compute the axial anomaly we will use Pauli-Villars regularization and
Mandelstam-Leibbrandt prescription to treat infrared divergences. We will
follow reference \cite{Pokorski}.

It is easy to check that formally:
\[ ( \Pi^{1 5\mu \nu \delta} + \Pi^{2 5\mu \nu \delta} + \Pi^{3 5\mu
   \nu \delta} + \Pi^{4 5\mu \nu \delta} ) p_{\delta} =0 \]
if shift of the integration variable $k \rightarrow k+Q$ is allowed. Here $Q$
is a constant vector.\footnote{This is true if we use DR as in chapter V and VI.}

Introduce a Pauli-Villars particle of mass $\bar{M}$ and define the
regularized amplitude:
\begin{eqnarray*}
  \Pi^{5R\mu \nu \delta} ( M, \bar{M} ,p,q ) = ( \Pi^{1 5\mu \nu \delta}
  + \Pi^{2 5\mu \nu \delta} + \Pi^{3 5\mu \nu \delta} + \Pi^{4 5\mu
  \nu \delta} ) ( M,p,q ) - ( \Pi^{1 5\mu \nu \delta} + \Pi^{2 5\mu \nu
  \delta} + \Pi^{3 5\mu \nu \delta} + \Pi^{4 5\mu \nu \delta} ) (
  \bar{M} ,p,q ) &  & 
\end{eqnarray*}
Since $\Pi^{5R\mu \nu \delta} ( M, \bar{M} ,p,q )$ is finite, it satisfies
the naive Ward identity(electric charge conservation):
\[ \Pi^{5R\mu \nu \delta} ( M, \bar{M} ,p,q ) p_{\delta} =0 \]
Besides, the axial Ward identity formally is,if shift of the integration variable $k \rightarrow k+Q$ is allowed:
\begin{eqnarray}
 - ( p+q )_{\mu} i ( \Pi^{1 5\mu \nu \delta} + \Pi^{2 5\mu \nu \delta}
   + \Pi^{3 5\mu \nu \delta} + \Pi^{4 5\mu \nu \delta} ) =2M\mathcal{A}
   ( M,p,q )^{\nu \delta} = \nonumber\\
   -2M ( -i e )^{2} \int d k  \{\tmop{Tr} \{ \gamma^{5}  \frac{i \left(
\left( \slashed{k} + \slashed{p} + \slashed{q} \right) +M- \frac{m^{2}}{2} 
\frac{\slashed{n}}{n \cdot ( k+p+q )} \right)}{( k+p+q )^{2} -M^{2} -m^{2} +i
\varepsilon} \left[ \gamma^{\nu} + \frac{1}{2} n^{\nu} \left( \slashed{n} \right)
m^{2}   ( n. ( k+p+q ) )^{-1} ( n. ( k+p ) )^{-1} \right] \nonumber\\
\frac{i \left(
\slashed{k} + \slashed{p} +M- \frac{m^{2}}{2}  \frac{\slashed{n}}{n \cdot ( k+p )}
\right)}{( k+p )^{2} -M^{2} -m^{2} +i \varepsilon} \left[ \gamma^{\delta} +
\frac{1}{2} n^{\delta} \left( \slashed{n} \right) m^{2}   ( n. ( k+p ) )^{-1} (
n.k )^{-1} \right] \frac{i \left( \slashed{k} +M- \frac{m^{2}}{2} 
\frac{\slashed{n}}{n \cdot ( k )} \right)}{( k )^{2} -M^{2} -m^{2} +i \varepsilon}\}
+ ( p, \delta ) \rightarrow ( q, \nu ) \}
\end{eqnarray}
\begin{eqnarray}
  ( -2M ) ( i e )^{2} n^{\delta} n^{\nu} i \int  d k ( n.k )^{-1} ( n. ( k-p-q
  ) )^{-1} [ ( n. ( k-q ) )^{-1} + ( n. ( k-p ) )^{-1} ]\nonumber\\
   \tmop{Tr} [\frac{1}{2}
  \slashed{n} m^{2}  \frac{i \left( \slashed{k} - \slashed{p} - \slashed{q} +M-
  \frac{m^{2}}{2}  \frac{\slashed{n}}{n \cdot ( k-p-q )} \right)}{( k-p-q )^{2}
  -M^{2} -m^{2} +i \varepsilon} \gamma^{5} &  &  \nonumber\\
  \frac{i \left( \slashed{k} +M- \frac{m^{2}}{2}  \frac{\slashed{n}}{n \cdot k}
  \right)}{k^{2} -M^{2} -m^{2} +i \varepsilon} ]  &  & 
  \label{4dano2}
\end{eqnarray}

The term (\ref{4dano2}) is convergent and has zero trace in four dimensions.
So it vanishes.

Therefore the regularized amplitude satisfies:
\[ - ( p+q )_{\mu} i \Pi^{5R\mu \nu \delta} ( M, \bar{M} ,p,q )
   =2M\mathcal{A} ( M,p,q )^{\nu \delta} -2 \bar{M} \mathcal{A} ( \bar{M} ,p,q
   )^{\nu \delta} \]
Since the original amplitude is obtained formally as $l i m_{\bar{M}
\rightarrow \infty}$, the axial anomaly is given by:
\[ A^{\nu \delta} =l i m_{\bar{M} \rightarrow \infty} (  -2\bar{M} \mathcal{A}
   ( \bar{M} ,p,q )^{\nu \delta} ) \]
After computing the trace, we use ML prescription to regulate the infrared
divergences.  $\mathcal{A} ( M,p,q )^{\nu \delta}$ is ultraviolet finite

A nice property of ML prescription is that preserve naive power counting.
Using this property, we can easily show that all terms containing $n^{\mu}$ in
$\mathcal{A} ( M,p,q )^{\nu \delta}$ are smaller than $M^{-2}$ for large $M$, so
they do not contribute to the axial anomaly.

$\mathcal{A}^{\nu \delta} = \lim_{\bar{M} \rightarrow \infty}$
\[ 8 \bar{M}^{2}  \varepsilon^{\nu \delta\alpha \beta }
   p_{\alpha} q_{\beta}  (-i e)^2  \int d^{4} k \frac{1}{( k+p )^{2} -m^{2} -
   \bar{M}^{2}} \frac{1}{k^{2} -m^{2} - \bar{M}^{2}} \frac{1}{( k+p )^{2}
   -m^{2} - \bar{M}^{2}} \]

That is:
\begin{eqnarray*}
  \mathcal{A}^{\nu \delta} =-(i e)^2 \frac{i}{2 \pi^{2}} \varepsilon^{\nu \delta \beta \mu} {p}_{\beta}
  {q}_{\mu}
\end{eqnarray*}
This is the standard result \cite{Pokorski}\cite{Peskin}.

We see that Pauli-Villars regularization of ultraviolet divergences and
Mandelstam-Leibbrandt regularization of infrared divergences preserve the Ward
identity for the vector current(electric charge conservation) as well as the
standard anomaly for the axial current, without modification from VSR terms.

\section{ Two dimensional axial anomaly in dimensional regularization}
To treat $\gamma^5$ we follow the prescription of \cite{mdr}. That is, in any number of
dimensions
\begin{eqnarray*}
	\gamma^{5} & =i  \gamma^{0} \gamma^{1} & \\
	\{ \gamma^{5} , \gamma^{\mu} \} =0. \mu =0,1; & [ \gamma^{5} , \gamma^{\mu}
	] =0, \mu =2,3 \ldots .,d & \\
	q_{\mu} ,n^{\mu}   \tmop{are}   \tmop{two}   \tmop{dimensional}  
	\tmop{vectors} . & p_{\mu}   \tmop{is}  d- \tmop{dimensional} & 
\end{eqnarray*}
\begin{eqnarray*}
	i  \Pi^{ 1 5\mu \nu} q_{\nu} =- ( -i e )^{2} \int d p  \tmop{Tr}
	\{ \left[ \gamma^{\mu} + \frac{1}{2} n^{\mu} \left( \slashed{n} \right)
	m^{2}   ( n. ( p+q ) )^{-1} ( n.p )^{-1} \right]  \frac{i \left( \slashed{p} +M-
		\frac{m^{2}}{2}  \frac{\slashed{n}}{n \cdot p} \right)}{p^{2} -M^{2} -m^{2} +i
		\varepsilon} \nonumber\\
	(\slashed{q} + \frac{1}{2} n.q  \left( \slashed{n} \right) m^{2} 
	( n. ( p+q ) )^{-1} ( n.p )^{-1} ) \gamma^{5} \frac{i \left( \left(
		\slashed{p} + \slashed{q} \right) +M- \frac{m^{2}}{2}  \frac{\slashed{n}}{n \cdot ( p+q
			)} \right)}{( p+q )^{2} -M^{2} -m^{2} +i \varepsilon} \} &  & 
\end{eqnarray*}
Write
\[ \slashed{p} = \slashed{p}_{1} + \slashed{p}_{2} ;1  \tmop{lives}   \tmop{in}  
\tmop{two}  \tmop{dimensions} ,2  \tmop{lives}   \tmop{in}  d-2  \tmop{dimensions} \]
Now we use the identity:
\begin{eqnarray}
	\left[ \slashed{q} + \frac{1}{2} n.q \left( \slashed{n} \right) m^{2}   ( n. ( p+q )
	)^{-1} ( n.p )^{-1} \right] =\nonumber\\
	\left[ \slashed{p} + \slashed{q} - \frac{1}{2} \slashed{n} m^{2} ( n. ( p+q )
	)^{-1} -M- \left( \slashed{p} - \frac{m^{2} \slashed{n}}{2n.p} -M \right) \right] & 
	& 
\end{eqnarray}

\begin{eqnarray}
	 \Pi^{1 5\mu \nu}q_{\nu} =- ( -i e )^{2} \int d p  \tmop{Tr} \{ - \left[ \gamma^{\mu} +
	\frac{1}{2} n^{\mu} \left( \slashed{n} \right) m^{2}   ( n. ( p+q ) )^{-1} ( n.p
	)^{-1} \right] \gamma_{5} \frac{i \left( \left( \slashed{p} + \slashed{q} \right)
		+M- \frac{m^{2}}{2}  \frac{\slashed{n}}{n \cdot ( p+q )} \right)}{( p+q )^{2}
		-M^{2} -m^{2} +i \varepsilon} + \nonumber\\
	\left[ \gamma^{\mu} + \frac{1}{2} n^{\mu} \left( \slashed{n} \right) m^{2}   (
	n. ( p+q ) )^{-1} ( n.p )^{-1} \right] 
	 \frac{i \left( \slashed{p} +M-
		\frac{m^{2}}{2}  \frac{\slashed{n}}{n \cdot p} \right)}{p^{2} -M^{2} -m^{2} +i
		\varepsilon} \gamma_{5} ( - ) \nonumber\\
	\left[ \left( \slashed{p}_{1} - \slashed{p}_{2} +
	\slashed{q} - \frac{1}{2} \slashed{n} m^{2} ( n. ( p+q ) )^{-1} -M \right) +2M
	\right] \frac{i \left( \left( \slashed{p} + \slashed{q} \right) +M- \frac{m^{2}}{2} 
		\frac{\slashed{n}}{n \cdot ( p+q )} \right)}{( p+q )^{2} -M^{2} -m^{2} +i
		\varepsilon} \} 
\end{eqnarray}
\begin{eqnarray}
	=- ( -i e )^{2} \int d p  \tmop{Tr} \left\{ \gamma^{5} \left[ \gamma^{\mu} +
	\frac{1}{2} n^{\mu} \left( \slashed{n} \right) m^{2}   ( n. ( p+q ) )^{-1} ( n.p
	)^{-1} \right] \frac{i \left( \left( \slashed{p} + \slashed{q} \right) +M-
		\frac{m^{2}}{2}  \frac{\slashed{n}}{n \cdot ( p+q )} \right)}{( p+q )^{2} -M^{2}
		-m^{2} +i \varepsilon} - \right.\label{1}\\
	\gamma^{5} \left[ \gamma^{\mu} + \frac{1}{2} n^{\mu} \left( \slashed{n} \right)
	m^{2}   ( n. ( p+q ) )^{-1} ( n.p )^{-1} \right] \frac{i \left( \slashed{p} +M-
		\frac{m^{2}}{2}  \frac{\slashed{n}}{n \cdot p} \right)}{p^{2} -M^{2} -m^{2} +i
		\varepsilon} \} \nobracket +\label{2}\\
	2M ( -i e )^{2} \int d p  \tmop{Tr} \left\{ \left[ \gamma^{\mu} +
	\frac{1}{2} n^{\mu} \left( \slashed{n} \right) m^{2}   ( n. ( p+q ) )^{-1} ( n.p
	)^{-1} \right]  \frac{i \left( \slashed{p} +M- \frac{m^{2}}{2}  \frac{\slashed{n}}{n
			\cdot p} \right)}{p^{2} -M^{2} -m^{2} +i \varepsilon} \gamma_{5} \frac{i
		\left( \left( \slashed{p} + \slashed{q} \right) +M- \frac{m^{2}}{2} 
		\frac{\slashed{n}}{n \cdot ( p+q )} \right)}{( p+q )^{2} -M^{2} -m^{2} +i
		\varepsilon} \right\}\nonumber\\
	- ( -i e )^{2} \int d p  \tmop{Tr} \left\{ \left[ \gamma^{\mu} + \frac{1}{2}
	n^{\mu} \left( \slashed{n} \right) m^{2}   ( n. ( p+q ) )^{-1} ( n.p )^{-1}
	\right]  \frac{i \left( \slashed{p} +M- \frac{m^{2}}{2}  \frac{\slashed{n}}{n \cdot
			p} \right)}{p^{2} -M^{2} -m^{2} +i \varepsilon} \gamma^{5 } 2 \slashed{p}_{2}
	\frac{i \left( \left( \slashed{p} + \slashed{q} \right) +M- \frac{m^{2}}{2} 
		\frac{\slashed{n}}{n \cdot ( p+q )} \right)}{( p+q )^{2} -M^{2} -m^{2} +i
		\varepsilon} \right\}\nonumber
\end{eqnarray}
In dimensional regularization we can shift variable $p \rightarrow p-q$ in the
term (\ref{1}). Then the addition of terms (\ref{1})  and (\ref{2} )is cancelled by the
contribution of Figure 2.

The anomaly is:
\begin{eqnarray}
	A^{\mu} =- ( -i e )^{2} \int d p  \tmop{Tr} \left\{ \left[ \gamma^{\mu} +
	\frac{1}{2} n^{\mu} \left( \slashed{n} \right) m^{2}   ( n. ( p+q ) )^{-1} ( n.p
	)^{-1} \right]  \frac{i \left( \slashed{p} +M- \frac{m^{2}}{2}  \frac{\slashed{n}}{n
			\cdot p} \right)}{p^{2} -M^{2} -m^{2} +i \varepsilon} \gamma_{5} 2
	\slashed{p}_{2} \frac{i \left( \left( \slashed{p} + \slashed{q} \right) +M-
		\frac{m^{2}}{2}  \frac{\slashed{n}}{n \cdot ( p+q )} \right)}{( p+q )^{2} -M^{2}
		-m^{2} +i \varepsilon} \right\} &  & 
\end{eqnarray}
That is:
\begin{eqnarray*}
	A^{\mu} =4 ( -i e )^{2} \int d p  \frac{\left[ -p_{2}^{2} \varepsilon^{\mu
			\nu} q_{\nu} -p_{2}^{2} \frac{1}{2} n^{\mu} m^{2}   ( n. ( p+q ) )^{-1} (
		n.p )^{-1} \varepsilon^{\alpha \beta} n_{\alpha} q_{\beta} \right] }{( p^{2}
		-M^{2} -m^{2} +i \varepsilon ) ( ( p+q )^{2} -M^{2} -m^{2} +i \varepsilon )}
	&  & 
\end{eqnarray*}
$p_{2}^{2} \sim ( d-2 ) p^{2}$ when $d \rightarrow 2$. The VSR part of the
integral is convergent, using ML prescription, so it is zero, when we take $d=2$. 

So only the Lorentz invariant part of the integral contributes to the anomaly.
\begin{eqnarray}
A^{\mu} =4e^{2} \varepsilon^{\mu \nu} q_{\nu} \int d p 
\frac{p_{2}^{2}}{p^{2} -M^{2} -m^{2} +i \varepsilon} \frac{}{( p+q )^{2}
	-M^{2} -m^{2} +i \varepsilon}=\nonumber\\
e^{2} \varepsilon^{\mu
	\nu} q_{\nu} \frac{i}{\pi}
\end{eqnarray}

That is:
\[ q_{\mu} <j^{5 \mu} ( q ) > = \frac{e^{2}}{-i e} \varepsilon^{\mu \nu}
q_{\nu} \frac{i}{\pi} A_{\mu} =- \frac{e}{\pi} \varepsilon^{\mu \nu}
q_{\nu} A_{\mu} = \frac{e}{\pi} \varepsilon^{\nu \mu} q_{\nu} A_{\mu} \]

which is the standard result\cite{Peskin}.

In Appendix B we study the vector Ward identity. If we use dimensional regularization there, then shifting the integration variable $p->p+Q$ is allowed.
So the naive Ward identity for the vector current is satisfied without anomaly.

\section{4d axial anomaly. Dimensional regularization}
In this section we compute the axial anomaly using dimensional regularization.
The contribution of Figure 3 is:
\begin{eqnarray*}
	- ( p+q )_{\mu} i  \Pi_{}^{15 \mu \nu \delta} =- ( -i e )^{2} \int d k 
	\tmop{Tr} \{ \left[ - \left( \slashed{p} + \slashed{q} \right) - \frac{1}{2} (
	p+q ) .n \left( \slashed{n} \right) m^{2}   ( n. ( k+q ) )^{-1} ( n. ( k
	-p ) )^{-1} \right] \gamma^{5} \\
	 \frac{i \left( \left( \slashed{k} +
		\slashed{q} \right) +M- \frac{m^{2}}{2}  \frac{\slashed{n}}{n \cdot ( k+q )}
		\right)}{( k+q )^{2} -M^{2} -m^{2} +i \varepsilon} \left[ \gamma^{\nu} +
	\frac{1}{2} n^{\nu} \left( \slashed{n} \right) m^{2}   ( n. ( k+q ) )^{-1} ( n.k
	)^{-1} \right] \frac{i \left( \slashed{k} +M- \frac{m^{2}}{2}  \frac{\slashed{n}}{n
			\cdot k} \right)}{k^{2} -M^{2} -m^{2} +i \varepsilon} \\
		\left[ \gamma^{\delta}
	+ \frac{1}{2} n^{\delta} \left( \slashed{n} \right) m^{2}   ( n. ( k-p ) )^{-1}
	( n.k )^{-1} \right] \frac{i \left( \slashed{k} - \slashed{p} +M- \frac{m^{2}}{2} 
		\frac{\slashed{n}}{n \cdot ( k-p )} \right)}{( k-p )^{2} -M^{2} -m^{2} +i
		\varepsilon} \} &  & 
\end{eqnarray*}

Write $\slashed{k} = \slashed{k}_{1} + \slashed{k}_{2}$
\begin{eqnarray*}
	\left( \slashed{k}_{1} + \slashed{k}_{2} + \slashed{p} + \slashed{q} +M \right) \gamma_{5}
	=- \gamma_{5} \left( \slashed{k}_{1} + \slashed{k}_{2} + \slashed{p} + \slashed{q} +M
	\right) +2 \gamma_{5} \slashed{k}_{2} +2M \gamma_{5} &  & 
\end{eqnarray*}

That is the anomaly is:
\begin{eqnarray}
\Gamma^{5\nu\delta}(p,q)=2 ( -i e )^{2} \int d k  \tmop{Tr} \{ \gamma^{5} \slashed{k}_{2}  \frac{i
	\left( \left( \slashed{k} + \slashed{p} + \slashed{q} \right) +M- \frac{m^{2}}{2} 
	\frac{\slashed{n}}{n \cdot ( k+p+q )} \right)}{( k+p+q )^{2} -M^{2} -m^{2} +i
	\varepsilon} \left[ \gamma^{\nu} + \frac{1}{2} n^{\nu} \left( \slashed{n} \right)
m^{2}   ( n. ( k+p+q ) )^{-1} ( n. ( k+p ) )^{-1} \right] \nonumber\\
\frac{i \left(
	\slashed{k} + \slashed{p} +M- \frac{m^{2}}{2}  \frac{\slashed{n}}{n \cdot ( k+p )}
	\right)}{( k+p )^{2} -M^{2} -m^{2} +i \varepsilon} \left[ \gamma^{\delta} +
\frac{1}{2} n^{\delta} \left( \slashed{n} \right) m^{2}   ( n. ( k+p ) )^{-1} (
n.k )^{-1} \right] \frac{i \left( \slashed{k} +M- \frac{m^{2}}{2} 
	\frac{\slashed{n}}{n \cdot ( k )} \right)}{( k )^{2} -M^{2} -m^{2} +i \varepsilon}
\}
\end{eqnarray}

To compute the trace, we notice that there must be an even number of
$\slashed{k}_{2}$ otherwise the trace vanishes. 
Assume there are four $\slashed{k}_{2}$
\begin{eqnarray*}
	\tmop{Tr} \left\{ \gamma^{5} \slashed{k}_{2} \slashed{k}_{2} \gamma^{\nu}
	\slashed{k}_{2} \gamma^{\delta} \slashed{k}_{2} \right\} = ( k_{2}^{2} )^{2}
	\tmop{Tr} \{ \gamma^{5} \gamma^{\nu} \gamma^{\delta} \} =0 &  & 
\end{eqnarray*}

That is, only two $\slashed{k}_{2}$ contribute to the trace.

The trace can be written as $\tmop{Tr} =k_{2}^{2} S$

But $k_{2}^{2} S \sim ( d-4 ) k^{2} S$. So if $k^{2} S$ is convergent in $d=4$
the contribution of this $S$ vanishes.If we use ML prescription to regularize the infrared divergences  we can show that $k^{2} S$ is convergent
in $d=4$ for all VSR $S$'s, since ML preserves naive power counting. Therefore only Lorentz invariant terms contribute to the anomaly.

Finally the anomaly is:
\begin{eqnarray}
\Gamma^{5\nu\delta}(p,q)=2 ( -i e )^{2} i^{3} \int d k k_{2}^{2}  \frac{\tmop{Tr} \left\{ \gamma_{5}
	\left( \slashed{q} \right) \gamma_{\nu} \left( \slashed{p} \right) \gamma_{\delta}
	\right\}}{( k+p+q )^{2} -M^{2} -m^{2} +i \varepsilon} \frac{1}{( k+p )^{2}
	-M^{2} -m^{2} +i \varepsilon} \frac{1}{( k )^{2} -M^{2} -m^{2} +i
	\varepsilon}\nonumber\\
=-i \frac{e^{2}}{2 \pi^{2}} \varepsilon^{\nu \delta \alpha \mu} p_{\alpha}
q_{\mu}
\end{eqnarray}

Therefore
\begin{eqnarray}
<p,q | \partial_{\mu} j^{5 \mu} ( 0 ) | 0> =- \frac{e^{2}}{2 \pi^{2}} \varepsilon^{\mu \nu \alpha \delta} ( -i q_{\mu} )
\varepsilon^{\ast}_{\nu} ( q ) ( -i p_{\alpha} ) \varepsilon^{\ast}_{\delta} (
p )
\end{eqnarray}

which is the standard result\cite{Peskin}.

Following the same reasoning as in  Appendix B, we can study the vector Ward identity in four dimensions. If we use dimensional regularization there, then shifting the integration variable $k->k+Q$ is allowed.
So the naive Ward identity for the vector current is satisfied without anomaly.

\section{Path integral derivation of the axial anomaly}

We use the approach of \cite{Fujikawa}.

The generating functional in the presence of an external field $A_\mu$ is;
\[ Z= \int \mathcal{D} \psi \mathcal{D} \bar{\psi}  e^{i \int d^{4} x
	\bar{\psi} i \slashed{\mathcal{D}} \psi} \]
where the gauge invariant and $Sim(2)$ invariant Dirac operator is

$\slashed{\mathcal{D}} =\slashed D + \frac{1}{2} \slashed{n} m^{2}  (n \cdot D)^{-1} ,
D_{\mu} = \partial_{\mu} -i e A_{\mu}$

Introduce a basis of eigenvectors of $\slashed{\mathcal{D}}$
\begin{eqnarray}
 i \slashed{\mathcal{D}} \phi_{m} = \lambda_{m} \phi_{m}  ,
\hat{\phi}_{m} \left( i \slashed{\mathcal{D}} \right) =-i \mathcal{D}_{\mu}
\hat{\phi}_{m} \gamma^{\mu} = \lambda_{m} \hat{\phi}_{m} 
\end{eqnarray}

For large $q$ and fixed $A_{\mu}$
\begin{eqnarray*}
	\phi_{m} ( x ) \sim \phi_{m} ( q ) e^{i q x} , &  & \lambda_{m}^{2} \sim
	q^{2} ,
\end{eqnarray*}

We can expand

\begin{eqnarray*}
	\psi ( x ) = \sum_{m} a_{m} \phi_{m} ( x ) , & \bar{\psi} ( x ) = \sum_{m}
	\bar{a}_{m} \hat{\phi}_{m} ( x ) & 
\end{eqnarray*}
The integration measure is defined by:
\[ \mathcal{D} \psi \mathcal{D} \bar{\psi} = \prod_{m} d a_{m} d  \bar{a}_{m}
\]
Under the change of variables:
\[ \psi' ( x ) = ( 1+i \alpha ( x ) \gamma^{5} ) \psi ( x ) \]
we get:
\begin{eqnarray}
	\mathcal{D} \psi' \mathcal{D} \bar{\psi}' =\mathcal{J}^{-2} \mathcal{D} \psi
	\mathcal{D} \bar{\psi} &  & 
\end{eqnarray}
where the jacobian $\mathcal{J}$ is given by:
\begin{eqnarray}
	\log \mathcal{J}=i \int d^{4} x \alpha ( x ) \sum_{n} \phi^{\dagger}_{n} (
	x ) \gamma^{5} \phi_{n} ( x )   &  & 
\end{eqnarray}

To evaluate it we introduce a gauge invariant \ and $\tmop{Sim} ( 2 )$
invariant regularization:
\begin{eqnarray*}
	\sum_{n} \phi^{\dagger}_{n} ( x ) \gamma^{5} \phi_{n} ( x )  = \lim_{M
		\rightarrow \infty} \sum_{n} \phi^{\dagger}_{n} ( x ) \gamma^{5} \phi_{n} (
	x ) e^{\frac{\lambda_{n}^{2}}{M^{2}}} &  & =\\
	\lim_{M \rightarrow \infty} <x \left| \tmop{Tr} \left\{ \gamma^{5}
	e^{\frac{\left( i \slashed{\mathcal{D}} \right)^2}{M^{2}}} \right\} \right| x>
	&  & 
\end{eqnarray*}
$\tmop{Tr}$ traces over Dirac indices.

We can write:
\[ \left( i \slashed{\mathcal{D}} \right)^{2} =-\mathcal{D}_{\mu}
\mathcal{D}_{\nu} g^{\mu \nu} - \frac{1}{2} [ \mathcal{D}_{\mu}
,\mathcal{D}_{\nu} ] \sigma^{\mu \nu} , \sigma^{\mu \nu} = \frac{i}{2} [
\gamma^{\mu} , \gamma^{\nu} ] \]
Since we take $M \rightarrow \infty$, we look at the asymptotic part of the
spectrum.

It is simpler to  evaluate the commutator in the light cone gauge $n.A=0$
\begin{eqnarray*}
	[ \mathcal{D}_{\mu} ,\mathcal{D}_{\nu} ] \phi \sim -i e F_{\mu \nu} ( p )
	\phi ( q ) e^{i ( p+q ) x} -i e \frac{1}{2} m^{2} \frac{n.p}{n.q ( n.p+n.q
		)} i \phi ( q ) e^{i ( p+q ) x} ( n_{\mu} A_{\nu} ( p ) -n_{\nu}  A_{\mu} (
	p ) ) &  & \\
	\sim -i e F_{\mu \nu} ( p ) \phi ( q ) e^{i ( p+q ) x}   \tmop{for large}  q & 
	& 
\end{eqnarray*}
\[ \left( i \slashed{\mathcal{D}} \right)^{2} =-\mathcal{D}_{\mu}
\mathcal{D}_{\nu} g^{\mu \nu} + \frac{e}{2} F_{\mu \nu} \sigma^{\mu \nu} \]
\begin{eqnarray*}
	\lim_{M \rightarrow \infty} <x \left| \tmop{Tr} \left\{ \gamma^{5}
	e^{\frac{\left( i \slashed{\mathcal{D}}^{2} \right)}{M^{2}}} \right\} \right|
	x>= &  & \\
	\lim_{M \rightarrow \infty} \tmop{Tr} \left\{ \frac{1}{2!} \gamma^{5} \left(
	\frac{e}{2M^{2}} F_{\mu \nu} \sigma^{\mu \nu} \right)^{2} \right\} <x \left|
	e^{- \frac{\partial^{2}}{M^{2}}} \right| x> &  & 
\end{eqnarray*}

\begin{eqnarray*}
	<x \left| e^{- \frac{\partial^{2}}{M^{2}}} \right| x>= \lim_{x \rightarrow
		y} \int \frac{d^{4} k}{( 2 \pi )^{4}} e^{-i k ( x-y )}
	e^{\frac{k^{2}}{M^{2}}} & = & \\
	i  \int \frac{d^{4} k_{E}}{( 2 \pi )^{4}} e^{- \frac{k^{2}_{E}}{M^{2}}} =i
	\frac{M^{4}}{16 \pi^{2}} &  & 
\end{eqnarray*}
Then:
\begin{eqnarray*}
	\lim_{M \rightarrow \infty} <x \left| \tmop{Tr} \left\{ \gamma^{5}
	e^{\frac{\left( i \slashed{\mathcal{D}}^{2} \right)}{M^{2}}} \right\} \right|
	x>=- \frac{e^{2}}{32 \pi^{2}} \varepsilon^{\alpha \beta \mu \nu}
	F_{\alpha \beta} ( x ) F_{\mu \nu} ( x ) &  & 
\end{eqnarray*}
That is:
\[ \mathcal{J}= \exp \left( -i \int d^{4} x \alpha ( x ) \frac{e^{2}}{16
	\pi^{2}} \varepsilon^{\alpha \beta \mu \nu} F_{\alpha \beta} ( x
) F_{\mu \nu} ( x ) \right) \]
Then the Adler-Bell-Jackiw anomaly follows.

Notice that we could get this result assuming that the infrared regulator of
$\frac{1}{n.q}$ preserves scaling(naive power counting). To garanty this
property we work with the ML prescription, as in the perturbative approach.

\section{Conclusions}

We have examined the appearance of axial anomalies in VSR electrodynamics, using Pauli-Villars and dimensional
regularization of ultraviolet divergences and Mandelstam-Leibbrandt regularization of infrared divergences.

Given that ML preserves naive power counting in loop integrals, we have shown that the usual form for the anomaly of the axial current
appears, without corrections from VSR terms. No anomaly is present  in the vector current conservation. This computation is at variance from a previous result for the axial anomaly in two dimensions \cite{as},where corrections from VSR terms were found. This difference could be due to different normalization conditions for the anomaly term\cite{Pokorski}  or some extra freedom that occurs when Lorentz invariance is violated\cite{aas}. In any case, our result implies that the procedure of \cite{as}
destroys the naive power counting of loop integrals.

In four dimension we find a completely different result compared to \cite{alex}. There they claim that the conservation for the vector current has an anomaly and VSR corrections should appear in the anomaly of the axial current. We notice also that Figure 5,6 are lacking in the computation of both anomalies in \cite{alex}. Figure 5,6 are crucial to satisfy the Ward identity for the vector current as a procedure in 4d similar to the one explained in Appendix B shows.

We study also the axial anomaly from the point of view of the path integral method. Again ML property of preserving scaling(naive power counting) permits to show that
the axial anomaly is the Lorentz invariant one, without corrections from VSR.

Finally we want to recall that $M$ is not the mass of the particle. So if the fermion acquires a VSR mass $m$ even if $M=0$, the divergence of the axial current will contain the anomalous term only.

\section{Acknowledgements}

The research of J.A. was partially supported by the Institute of Physics at
PUC.

\section*{Appendix A:Feynman rules}

\begin{figure}[h!]
	\centering
	\includegraphics[scale=0.65]{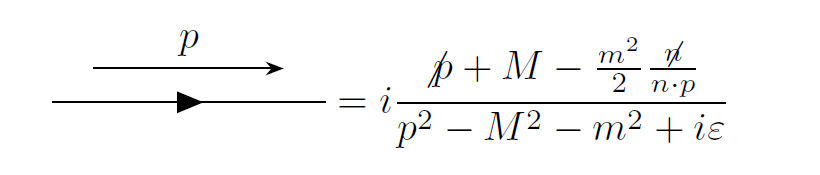}
	\caption{Electron propagator}
	\label{fig:ep}
\end{figure}

\begin{figure}[h!]
	\centering
	\includegraphics[scale=0.65]{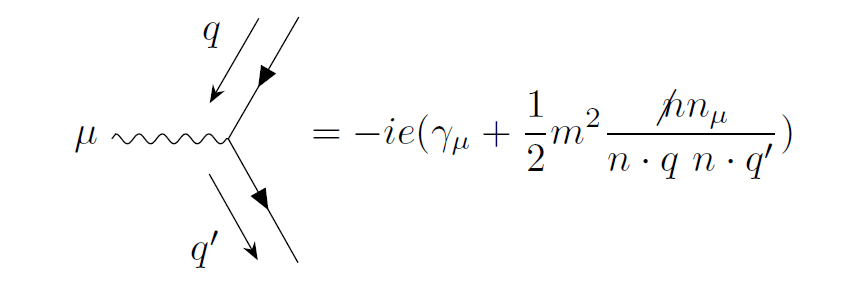}
	\caption{$e-e-A_{\mu}$
		vertex}
	\label{fig:v3}
\end{figure}
\begin{figure}[h!]
	\centering
	\includegraphics[scale=0.65]{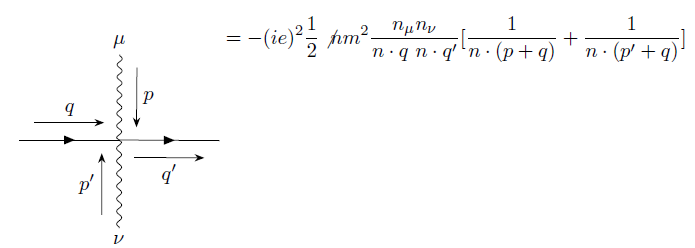}
	\caption{$e-e-A_{\mu}
		-A_{\nu}$ vertex}
	\label{fig:ep}
\end{figure}
\begin{figure}[h!]
	\centering
	\includegraphics[scale=0.65]{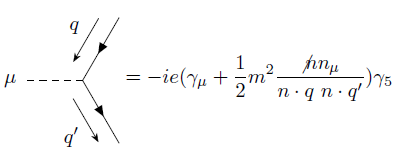}
	\caption{axial-e-e
		vertex}
	\label{fig:av3}
\end{figure}
\begin{figure}[h!]
	\centering
	\includegraphics[scale=0.65]{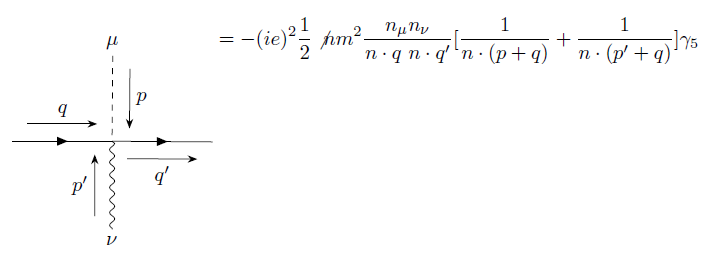}
	\caption{$axial
		-A_{\nu} -e-e$ vertex}
	\label{fig:av4}
\end{figure}
\begin{figure}[h!]
	\centering
	\includegraphics[scale=0.65]{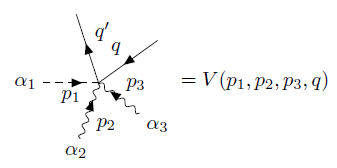}
	\caption{$\tmop{axial}
		-A_{\alpha_{2}} -A_{\alpha_{3}} -e-e$ vertex}
	\label{fig:aV}
\end{figure}

\begin{eqnarray*}
  V ( p_{1} ,p_{2} ,p_{3} ,q ) =i (ie)^{3} \frac{m^{2}}{2} \slashed{n}
  n^{\alpha_{1}} n^{\alpha_{2}} n^{\alpha_{3}}  \frac{1}{n. (q+p_{1} +p_{2}
  +p_{3} )}\\
  ( \frac{1}{n. (q+p_{1} +p_{2} )}  \frac{1}{n. (q+p_{1} )} + \frac{1}{n.
  (q+p_{1} +p_{2} )}  \frac{1}{n. (q+p_{2} )} +\\
  \frac{1}{n. (q+p_{3} +p_{2} )}  \frac{1}{n. (q+p_{3} )} + \frac{1}{n.
  (q+p_{1} +p_{3} )}  \frac{1}{n. (q+p_{1} )} +\\
  \frac{1}{n. (q+p_{2} +p_{3} )}  \frac{1}{n. (q+p_{2} )} + \frac{1}{n.
  (q+p_{3} +p_{1} )}  \frac{1}{n. (q+p_{3} )} ) \gamma^{5}
\end{eqnarray*}

\section*{Appendix B:Formal proof of the Ward identities in 2d}

In this appendix we want to show in some detail how to obtain the Ward
identities in 2d. In 4d we have more graphs, but the procedure is essentially
the same.
\begin{eqnarray*}
  q_{\mu} \Pi^{1 5\mu \nu} =- ( -i e )^{2} \int d p  \tmop{Tr} \{
  \left[ \slashed{q} + \frac{1}{2} n.q \left( \slashed{n} \right) m^{2}   ( n. ( p+q )
  )^{-1} ( n.p )^{-1} \right]  \frac{i \left( \slashed{p} +M- \frac{m^{2}}{2} 
  \frac{\slashed{n}}{n \cdot p} \right)}{p^{2} -M^{2} -m^{2} +i \varepsilon}\nonumber\\
  \left[ \gamma^{\nu} + \frac{1}{2} n^{\nu} \left( \slashed{n} \right) m^{2}   (
  n. ( p+q ) )^{-1} ( n.p )^{-1} \right] \gamma^{5} \frac{i \left( \left(
  \slashed{p} + \slashed{q} \right) +M- \frac{m^{2}}{2}  \frac{\slashed{n}}{n \cdot ( p+q
  )} \right)}{( p+q )^{2} -M^{2} -m^{2} +i \varepsilon} \} &  & 
\end{eqnarray*}
Now we use the identity:
\begin{eqnarray}
  \left[ \slashed{q} + \frac{1}{2} n.q \left( \slashed{n} \right) m^{2}   ( n. ( p+q )
  )^{-1} ( n.p )^{-1} \right] =\nonumber\\
   \left[ \slashed{p} + \slashed{q} - \frac{1}{2} \slashed{n} m^{2} ( n. ( p+q )
  )^{-1} -M- \left( \slashed{p} - \frac{m^{2} \slashed{n}}{2n.p} -M \right) \right] & 
  & 
\end{eqnarray}
and the cyclic property of the trace to get:
\begin{eqnarray}
  q_{\mu} \Pi^{1 5\mu \nu} = ( -i e )^{2} \int d p  \tmop{Tr} \{
  \frac{ \left( \slashed{p} +M- \frac{m^{2}}{2}  \frac{\slashed{n}}{n \cdot p}
  \right)}{p^{2} -M^{2} -m^{2} +i \varepsilon} 
\left[ \gamma^{\nu} +
  \frac{1}{2} n^{\nu} \left( \slashed{n} \right) m^{2}   ( n. ( p+q ) )^{-1} ( n.p
  )^{-1} \right] \gamma^{5} - \nonumber\\
  \left[ \gamma^{\nu} + \frac{1}{2} n^{\nu} \left(
  \slashed{n} \right) m^{2}   ( n. ( p+q ) )^{-1} ( n.p )^{-1} \right] \gamma^{5}
  \frac{ \left( \left( \slashed{p} + \slashed{q} \right) +M- \frac{m^{2}}{2} 
  \frac{\slashed{n}}{n \cdot ( p+q )} \right)}{( p+q )^{2} -M^{2} -m^{2} +i
  \varepsilon} \} &  & \label{2dwi1}
\end{eqnarray}
Besides:
\begin{eqnarray*}
  q_{\mu} \Pi^{2 5\mu \nu} = &  & \\
  {\color{red} } 2 ( i e )^{2} n.q n^{\nu} \int  d p ( n.p )^{-1} [ ( n. ( q+p
  ) )^{-1} ( n. ( -q+p ) )^{-1} ] \tmop{Tr} \frac{1}{2} \slashed{n} m^{2}  \frac{
  \left( \slashed{p} +M- \frac{m^{2}}{2}  \frac{\slashed{n}}{n \cdot p} \right)}{p^{2}
  -M^{2} -m^{2} +i \varepsilon} \gamma^{5} &  & 
\end{eqnarray*}

In the second term of(\ref{2dwi1}) shift $p \rightarrow p-q$ to get:
\footnote{This is justified if we use DR as in chapter V and VI.}

\begin{eqnarray*}
  q_{\mu} \Pi^{1 5\mu \nu} = &  & \\
  - ( -i e )^{2} \int d p  \tmop{Tr} \left\{ \frac{1}{2} n^{\nu} \left(
  \slashed{n} \right) m^{2} \frac{ \left( \slashed{p} +M- \frac{m^{2}}{2} 
  \frac{\slashed{n}}{n \cdot p} \right)}{p^{2} -M^{2} -m^{2} +i \varepsilon}
  \gamma^{5} ( n.p )^{-1} ( -2n.q ) [   ( n. ( p+q ) )^{-1} ( n. ( p-q )
  )^{-1} ] \right\} &  & 
\end{eqnarray*}
That is $\Pi^{\mu \nu} = \Pi^{1 \mu \nu} + \Pi^{2 \mu \nu}$ is transverse.

The axial Ward identity is obtained in the same way.

\end{document}